\documentclass[aps,prc]{revtex4-2}
\usepackage{graphicx}
\usepackage{amsmath}
\usepackage{amssymb}

\newcommand{\be}{\begin{equation}}
\newcommand{\ee}{\end{equation}}
\newcommand{\ba}{\begin{eqnarray}}
\newcommand{\ea}{\end{eqnarray}}
\newcommand{\ban}{\begin{eqnarray*}}
\newcommand{\ean}{\end{eqnarray*}}

\begin{document}

\title{Relativistic Effects in Femtoscopy and Deuteron Formation}

\author{Stanis\l aw Mr\' owczy\' nski} 
\affiliation{National Centre for Nuclear Research, ul. Pasteura 7,  PL-02-093 Warsaw, Poland}

\date{September 12, 2026}

\begin{abstract}

For a long time studies of femtoscopic correlations have provided information about space-time characteristics of particle sources in high-energy collisions. Recently, the correlation functions have been also used to determine interaction parameters of correlated particles which is especially important for short-lived particles, for which scattering experiment are impossible. The abundance of experimental data and their high accuracy require an improved theoretical approach to femtoscopic correlations. The aim of this paper is to systematically discuss relativistic effects and their role in femtoscopy. Since a general relativistic approach is currently unavailable, due to serious theoretical difficulties, the correlation functions must be computed in the center-of-mass frame where the correlated particles are mostly non-relativistic. This requires transforming the source function to this frame, the consequences of which are discussed. Since the deuteron formation has a similar physical origin to femtoscopic correlations, relativistic effects are also discussed in the former process. The considerations are illustrated with calculations of some correlation functions and the deuteron coalescence coefficient to demonstrate a magnitude of relativistic effects. 

\end{abstract}

\maketitle

\section{Introduction}

Femtoscopic correlations of two particles with small relative momenta are determined by the distribution of particle emission points, encoded in the source function, and their wave function which depends on the particle statistics and their mutual interaction. Therefore, by measuring the correlations of pairs of protons or of pions whose interactions are known, the correlation functions provide information about  space-time characteristics of particle sources in high-energy collisions. Such measurements have been conducted for years, see the review articles \cite{Wiedemann:1999qn,Heinz:1999rw,Lisa:2005dd}, but recently, thanks to great improvements in experimental methods, the problem has been reversed, as discussed in the review \cite{Fabbietti:2020bfg}. By measuring particle correlations of various particle species, it is possible to reliably assume that the source function of a given particle type is known, and femtoscopic correlations can be used to determine unknown interaction parameters. This is especially important in case of particles that have such a short lifetime that it is impossible to perform a typical scattering experiment. Needless to say, this is the case for most hadrons.

The abundance of experimental data and their high accuracy require an improved theoretical approach to femtoscopic correlation functions, which have traditionally been described applying a very successful though simplified Lednick\'y-Luboshitz method \cite{Lednicky:1981su,Gmitro:1986ay,Lednicky:2005tb}. The method uses the asymptotic form of the scattering  wave function and takes into account only the $s$-wave scattering amplitude. Recently, there have been some efforts to go beyond these limitations, see {\it e.g.} \cite{Gobel:2025afq,Murase:2025nlo,Romanenko:2026wkn}. The aim of this study is instead to discuss in some detail relativistic effects which have not been adequately addressed. 

Hadrons produced in high-energy collisions are typically relativistic, so the description of correlations of these hadrons should be relativistic. However, such a description is available only under some rather drastic simplifying assumptions. In particular, a relativistically covariant formalism is known for the case when the inter-particle interaction is ignored and the correlations are solely caused by quantum statistics \cite{Wiedemann:1999qn}. In general, a theory of strongly interacting relativistic particles is an unsolved problem of theoretical physics. 

Three fundamental difficulties need to be overcome. The first arises at the level of classical dynamics and involves the inability to separate the center-of-mass (CM) from relative motion, see e. g. \cite{Alba:2006hs} where a history of the problem is also briefly reviewed. The second concerns a quantum theory and is related to the fact that the Lorentz boost operator does not commute with the particle number operator. Consequently, the number of  system's constituents depends on the reference frame. The third difficulty is very general: there are no universal method to solve equations of motion of strongly interacting systems. 

The matter becomes much simpler if the interaction is weak enough that a perturbation calculus is applicable. Nevertheless, it should be remembered that the weakness of the interaction does not guarantee the applicability of perturbative methods. This is well known in case of electromagnetic interactions where scattering problems are solved with fantastic accuracy while a description of relativistic bound states, which is by its very nature nonperturbative, encounters fundamental difficulties only partially resolved. The whole issue is well explained and discussed in the monograph \cite{Hoyer:2021adf}.

\newpage

Fortunately, the problem of interest is greatly simplified by the following circumstance: sizable femtoscopic correlations occur only at small relative momenta of two particles, typically less than 100~${\rm MeV}/c$. Therefore, the particle relative motion can be treated as nonrelativistic in the CM reference frame. Only in case of pions -- the lightest hadrons -- relativistic effect of relative motion can be significant but they can be taken into account as corrections. The motion of femtoscopically correlated hadrons relative to the source can be truly relativistic and it requires a fully relativistic treatment. 

Such a situation suggests an approach in which the interaction of hadrons is described in their CM frame. Then, one can use the Schr\"odinger equation with a whole machinery of nonrelativistic quantum mechanics. The source function, however, needs to be Lorentz transformed to the CM frame. Fortunately, unlike the wave function, the source function has simple transformation properties due to its probabilistic nature. This is the approach, which is adopted in this paper, and I discuss how the correlation functions depend on the particle velocity relative to the source. It should be clearly stated that the approach is not original. It is commonly accepted that femtoscopic correlations should be studied in the CM frame of particle pairs. However, the need to Lorentz transform the source function and its consequences have not been fully recognized. 

For a long time femtoscopic measurements mainly concerned correlations of identical pions. Their strong interaction could be and was neglected while the Coulomb effect was treated as a correction which was subtracted from experimental data, using the Bowler-Sinyukov \cite{Bowler:1991vx,Sinyukov:1998fc} or similar procedure. The experimental data were therefore intended to represent the correlations of free pions. Since, as already mentioned, the correlations of non-interacting particles can be described in a relativistically covariant way and the Bowler-Sinyukov correction factor is a Lorentz scalar, the approach to the femtoscopic correlations is approximate but relativistically covariant. The procedure works well for pions but is less satisfactory for kaons \cite{Maj:2009ue}. 

Until recently the most advanced approach was that proposed by Lednick\'y and Luboshitz \cite{Lednicky:1981su,Gmitro:1986ay,Lednicky:2005tb}. These authors' starting point is the relativistically covariant Bethe-Salpeter equation, which suggests that the entire formalism is relativistically covariant. Howver, for the reasons described above, the authors work in the CM frame of correlated particles. Since solutions of the Bethe-Salpeter equation are known only in a very limited number of cases, the authors adopted the assumptions that in fact are satisfied in the nonrelativistic limit. These assumptions include the factorization of the relative and CM motions of two particles, as well as the vanishing of the so-called relative time. As a result, the Bethe-Salpeter amplitude can be replaced by a nonrelativistic wave function. Instead of starting with a relativistic approach, which becomes non-relativistic due to the assumptions made, I propose to use a non-relativistic approach in the CM frame from the beginning, perform a Lorentz transformation of the source function, and finally take advantage of the fact that the correlation function is a Lorentz scalar which can be easily transformed to any reference frame. 

In full analogy to the femtoscopic correlations one treats a deuteron production in the coalescence model \cite{Butler:1963pp,Schwarzschild:1963zz}, where the deuteron production is a two-step process: production of nucleons and formation of a deuteron due to final-state interaction of neutron and proton that are close in phase-space. The energy scale of the first step, which is of order of nucleon mass, is much bigger than that of the second one which is a deuteron binding energy. Consequently, a probability to produce a deuteron factorizes into the probabilities to produce a neutron-proton pair and to form the deuteron. The latter probability is given by the Sato-Yazaki formula \cite{Sato:1981ez}, which is fully analogous to the femtoscopic Koonin formula \cite{Koonin:1977fh}, as discussed in detail in \cite{Mrowczynski:1992gc}. The argument regarding the difference in energy scales characterizing the nucleon-pair production and deuteron formation applies when nucleons are actually produced, rather than being fragments of colliding nuclei. This situation occurs in high-energy collisions of leptons and hadrons. In case of collisions of atomic nuclei, the argument also works but only for nucleons that do not originate from the fragmentation of a projectile or target nucleus. So, I am interested in truly produced deuterons, as those at mid-rapidity in experiments at RHIC and the LHC. 

Since measurements of femtoscopic correlations are typically performed on relativistic particles, various aspects of relativistic effects have naturally been addressed in the literature. The aim of this paper is to systematically discuss these effects in a pedagogical manner. For this reason, the considerations presented in the introductory three short sections are not original; however, they provide essential background for understanding the problem, as well as for the discussion in Secs. 5 and 6, which present original results on the influence of relativistic effects on femtoscopic correlation functions and deuteron coalescence coefficients. It is also demonstrated that a simultaneous analysis of these two quantities allows for a more straightforward assessment of the role of relativistic effects.

The discussion of relativistic effects in femtoscopy begins with the definition of the two-particle correlation function, which clearly demonstrates that the function is a Lorentz scalar. However, to make use of this property, the correlation function cannot be averaged -- as is commonly done -- over the total momenta of particle pairs. In Sec.~\ref{sec-koonin}, I discuss the nonrelativistic Koonin formula \cite{Koonin:1977fh}, which expresses the correlation function in terms of the source function and the nonrelativistic wave function of the two particles of interest. A relativistic form of the source function and its transformation properties are presented in Sec.~\ref{sec-source-function}. In the subsequent section, the $\Lambda$–$p$ and $p$–$p$ correlation functions are computed to quantitatively assess the magnitude of relativistic effects. Specifically, I consider a particle source that is isotropic in its rest frame but becomes anisotropic due to the Lorentz transformation to the CM frame of the particle pair. The $\pi$–$\pi$ correlation functions are also discussed here. Sec.~\ref{sec-deuteron} is devoted to deuteron formation, where it is shown that relativistic effects are more pronounced in the coalescence coefficient than in the correlation function. When the source function of nucleons is inferred from the proton-proton or lambda-proton correlation function, the Sato-Yazaki formula \cite{Sato:1981ez} uniquely determines the deuteron coalescence coefficient. 
The paper concludes with a summary of the considerations and findings. In the Appendix, I rederive the wave function used in the Lednick\'y-Lyuboshitz method \cite{Lednicky:1981su,Gmitro:1986ay,Lednicky:2005tb}, which accounts for Coulomb repulsion together with $s$-wave scattering due to the strong interaction.

I frequently refer to experimental data, but only to those collected at RHIC and the LHC. This is done so for several reasons. I am primarily interested in relativistic effects, which are particularly important in collisions at the highest accessible energies. Furthermore, the theoretical description of the processes studied in collider experiments is simpler than that of experiments with fixed targets. The laboratory frame at RHIC and the LHC coincides with the CM frame of the colliding identical nuclei and therefore also with the CM frame of the particle source. Moreover, the data are collected predominantly in the narrow mid-rapidity region. When the rapidity of a given particle vanishes, its transverse momentum equals its total momentum. In the context of deuteron production, it is also important that deuterons observed at mid-rapidity are not fragments of the incoming nuclei but are genuinely produced in the collision. It is also worth noting that relatively recent publications of results from RHIC and the LHC typically contain extensive references to earlier work.

Having said all this, it should be emphasized that femtoscopic correlations and deuteron production have been studied for years in nuclear collisions at low collision energies (tens of MeV) and intermediate energies (hundreds of MeV), see the review  \cite{Verde:2006dh} and recent article \cite{Si:2025eou}. Numerous new methods have been developed, yielding many interesting results. Because both femtoscopic correlations and deuteron formation involve the interactions between particles with low relative velocities, the energy of the collisions in which these particles are produced is not the determining factor. Therefore, the experimental methods and theoretical approaches to these processes at low and high energies are similar. However, I am primarily interested in the effect of particle motion relative to the particle source, which in the case of nuclear collisions at low and intermediate energies is approximately nonrelativistic. As a result, the effects considered in this work are small, if not entirely negligible, and thus such collisions are not discussed. 

Throughout the paper I use natural units, where $c = \hbar = 1$, and the metric convention is $(+,-,-,-)$.

\section{Definition of correlation function}
\label{sec-def}

The correlation function $C(\mathbf{p}_{1}, \mathbf{p}_{2})$ of two particles $a$ and $b$ with momenta $\mathbf{p}_{1}$ and $\mathbf{p}_{2}$ is defined as
\be
\label{def-corr-fun}
C({\bf p}_1, {\bf p}_2)
=\frac{\frac{dN_{ab}}{d{\bf p}_1 d{\bf p}_2}}
{\frac{dN_a}{d{\bf p}_1}\frac{dN_b}{d{\bf p}_2}} \;,
\ee
where $\frac{dN_{ab}}{d\mathbf{p}_1 d\mathbf{p}_2}$ and $\frac{dN_a}{d\mathbf{p}_1}$, $\frac{dN_b}{d\mathbf{p}_2}$ are, respectively, the two- and one-particle momentum distributions. The correlation function can be written down in the Lorentz covariant form
\be
\label{def-cov}
C({\bf p}_1, {\bf p}_2)=
\frac{E_1 E_2\frac{dN_{ab}}{d\mathbf{p}_1 d\mathbf{p}_2}}
{E_1\frac{dN_a}{d\mathbf{p}_1} E_2\frac{dN_b}{d\mathbf{p}_2}} \;,
\ee
where $E\frac{dN}{d^3\mathbf{p}}$ is the Lorentz invariant distribution.

The covariant form (\ref{def-cov}) shows that the correlation function is a Lorentz scalar field which can be easily transformed from one reference frame to another. Since the particle four-momenta are on mass-shell, the transformation of three-momenta ${\bf p}_i \rightarrow {\bf p'}_i$ with $i=1,2$ is well defined but it is not linear. It is natural to discuss the Lorentz transformation using the four-momenta: $p_1 =  (E_1,{\bf p}_1)$, $p_2 = (E_2,{\bf p}_2)$ with $E_1 = \sqrt{m_a^2 + {\bf p}_1^2}$, $E_2 = \sqrt{m_b^2 + {\bf p}_2^2}$. Since the four-momenta transform under the Lorentz transformation as $p_i  \rightarrow p'_i = L p_i$, where $L$ is the transformation matrix, the correlation function transforms as
\be
\label{corr-fun-transform}
C(p_1, p_2) \longrightarrow C'(p'_1,p'_2) = C (L^{-1}p'_1,L^{-1}p'_2) ,
\ee
where $L^{-1}$ denotes the inverse transformation matrix. 

Femtoscopic correlation functions are usually not measured as functions of three-momenta of both particles. Experimentally obtained correlation functions typically depend only on the difference of three-momenta of the two particles or the three-momentum at the center of mass of the two particles. Very often there are correlation functions measured only as a function of the length of the relative three-momentum. Then, there is no way to transform the correlation function to a reference frame other than that one in which the measurement was made.

To explain the problem I consider a correlation function of two identical particles measured as a function of ${\bf q} = \frac{1}{2}({\bf p}_1 - {\bf p}_2)$. Since the two particles are on mass-shell, the time-like component of the four-vector $Q^\mu = \frac{1}{2}(p_1^\mu - p_2^\mu)$ is $Q^0 = \frac{1}{2}(E_1 - E_2)$ and one shows that $Q^0 = {\bf q} \cdot {\bf v}$ where ${\bf v} = ({\bf p}_1 + {\bf p}_2)/(E_1 + E_2)$ is the velocity of the pair. So, if the correlation function is measured as function of ${\bf q}$ at a fixed value of the total momentum ${\bf P} = {\bf p}_1 + {\bf p}_2$, one can construct the four-vector $Q = ({\bf q} \cdot {\bf v}, {\bf q})$ and the correlation function can be then transformed as a scalar that is $C(Q) \longrightarrow C'(Q') = C (L^{-1}Q')$. However, correlation functions are usually measured not at fixed ${\bf P}$ but rather for a whole domain of ${\bf P}$. So, the particle pairs with different ${\bf v}$ contribute to the correlation function. Consequently, the three-vector ${\bf q}$ cannot be extended to the four-vector $Q$ and such a correlation function cannot be transformed. The situation is even worse when the correlation function is measured only as a function of $|{\bf q}|$. 

\section{Nonrelativistic Koonin Formula}
\label{sec-koonin}

Within the Koonin model \cite{Koonin:1977fh}, the correlation function is given as
\be 
\label{Koonin} 
C(\mathbf{p}_1,\mathbf{p}_2) = 
\int d^3r_1 dt_1 \int d^3r_2 dt_2 \, S(t_1,\mathbf{r}_1) \,
S(t_2,\mathbf{r}_2) \, |\Psi(\mathbf{r}'_1,\mathbf{r}'_2)|^2 , 
\ee
where $\mathbf{r}'_i \equiv \mathbf{r}_i+\mathbf{v}_i t_i$ with $i=1,2$, $\Psi(\mathbf{r}'_1,\mathbf{r}'_2)$ is the wave function of the two particles and $S(t,\mathbf{r})$ is the single-particle source function which gives the probability to emit a particle from the space-time point $(t,\mathbf{r})$. The source function is normalized as 
\be
\label{norma}
\int d^3r \, dt \, S(t,\mathbf{r})=1 .
\ee

After changing the variables ${\bf r}_i \rightarrow {\bf r}'_i$, the correlation function (\ref{Koonin}) can be written in the form
\be
C(\mathbf{p}_{1}, \mathbf{p}_{2}) = 
\int d^3 r_1 dt_1\int d^3 r_2 dt_2 \,
S(t_1,\mathbf{r}_1-\mathbf{v}_1t_1) \,
S(t_2,\mathbf{r}_2-\mathbf{v}_2t_2) \,
|\Psi(\mathbf{r}_1,\mathbf{r}_2)|^2 ,
\ee
where the primes are omitted.

Now, one introduces the CM coordinates 
\be
{\Large\Bigg\{}
\begin{array}{l}
\mathbf{r}=\mathbf{r}_2-\mathbf{r}_1, 
\\[2mm]
t=t_2-t_1, 
\\[2mm]
\mathbf{q}= \frac{1}{M}(m_2 \mathbf{p}_1 - m_1\mathbf{p}_2) ,
\end{array} 
~~~~~~~~~~~~
{\Large\Bigg\{}
\begin{array}{l}
\mathbf{R}=\frac{1}{M}(m_1\mathbf{r}_1+m_2\mathbf{r}_2), 
\\[2mm]
T=\frac{1}{M}(m_1 t_1+m_2 t_2),
\\[2mm]
\mathbf{P}=\mathbf{p}_1+\mathbf{p}_2,
\end{array}
\ee
where $M \equiv m_1+m_2$. Using the CM variables and factorizing the wave function as
\be
\Psi(\mathbf{r}_1,\mathbf{r}_2) =
e^{i\mathbf{P}\mathbf{R}}\varphi_\mathbf{q}(\mathbf{r}) ,
\ee
where $\varphi_\mathbf{q}(\mathbf{r})$ is the wave function of the relative motion, one gets
\be 
\label{Koonin-r-4D} 
C(\mathbf{q},\mathbf{P})= \int d^3r \, dt  \, S_r(t,\mathbf{r}-\mathbf{v}t)
|\varphi_\mathbf{q}(\mathbf{r})|^2 .
\ee
The relative source function $S_r(t,\mathbf{r})$ is expressed through the single-particle source function in the following way
\be
\label{source-relat}
S_r(t,\mathbf{r}) \equiv \int d^3R \, dT \,
S\Big(T-\frac{m_2}{M}t,\mathbf{R}-\frac{m_2}{M}\mathbf{r}\Big) \,
S\Big(T+\frac{m_1}{M}t,\mathbf{R}+\frac{m_1}{M}\mathbf{r}\Big) .
\ee
To get Eq.~(\ref{Koonin-r-4D}) it has been assumed that the velocity ${\bf v}$ is the same for both particles. Thus, one assumes that $\mathbf{v}_1=\mathbf{v}_2=\mathbf{v}$ which requires, strictly speaking, ${\bf q} = 0$. However, it is observed that $|{\bf v}_1 - {\bf v}_2 | \ll |{\bf v}_i|$ if $|{\bf q} | \ll \mu |{\bf p}_i|/m_i$ where $\mu \equiv m_1 m_2/M$. Thus, the approximation $\mathbf{v}_1 \approx \mathbf{v}_2$ holds for sufficiently small particle's momenta in the CM frame. It should be stressed that the dependence of the correlation function on ${\bf q}$ is mostly controlled by the dependence of the wave function on ${\bf q}$ which is not influenced by the above approximation.

One can introduce the effective relative source function which depends on ${\bf v}$ and is defined as 
\be
\label{effective-r-source}
S_r^{\rm eff}(\mathbf{r}) \equiv \int dt \, S_r(t,\mathbf{r-v}t) .
\ee
Then, the formula (\ref{Koonin-r-4D}) gets the simpler form
\be 
\label{Koonin-r-3D} 
C(\mathbf{q},\mathbf{P}) = \int d^3r\, S_r^{\rm eff} (\mathbf{r})
|\varphi_\mathbf{q}(\mathbf{r})|^2 . 
\ee
Due to the normalization condition (\ref{norma}), the functions $S_r(t,\mathbf{r})$ and $S_r^{\rm eff}(\mathbf{r})$ are also normalized
\be
\int d^3r \, S_r(\mathbf{r}) = \int d^3r \, dt\, S_r^{\rm eff}(t,\mathbf{r}) = 1 .
\ee

Since the correlation function depends, in general, on ${\bf p}_1$ and ${\bf p}_2$, it depends on ${\bf q}$ and ${\bf P}$ after the change of variables. The correlation function (\ref{Koonin-r-4D}) or (\ref{Koonin-r-3D}) depends on ${\bf P}$ through the velocity ${\bf v}$ of the particle pair. However, as already mentioned, femtoscopic correlations functions are usually measured only as functions of ${\bf q}$ not of ${\bf q}$ and ${\bf P}$. The dependence on ${\bf P}$ is usually averaged over. However, I will keep explicitly the dependence on ${\bf P}$ because, as explained in the introduction, the dominant relativistic effect enters through this dependence. 

To describe experimental data with the Koonin formula either in the form (\ref{Koonin}), (\ref{Koonin-r-4D}) or (\ref{Koonin-r-3D}), the source function $S(t,{\bf r})$ has to be determined. It is commonly used in the Gaussian parameterization 
\be
\label{gauss}
S(t,\mathbf{r})=\frac{1}{4\pi^2 R_x R_y R_z \tau} \:
\exp\Big[- \frac{t^2}{2\tau^2} -\frac{x^2}{2R_x^2} 
- \frac{y^2}{2R_y^2} - \frac{z^2}{2R_z^2}\Big],
\ee
where ${\bf r}=(x,y,z)$ and the parameters $\tau$, $R_x$, $R_y$ and $R_z$ characterize the lifetime and sizes of the source. Specifically, the parameters $\tau$ and $R_x$ give, respectively,
\be
\tau^2 = \langle t^2 \rangle \equiv \int d^3r\:dt\: t^2
S(t,\mathbf{r})
\;,\;\;\;\;\;\;\;\;\;
R_x^2 = \langle x^2\rangle \equiv \int d^3r \: dt\: x^2
S(t,\mathbf{r}) .
\ee

The relative source function computed from Eq.~(\ref{source-relat}) with the single-particle source (\ref{gauss}) is
\be
\label{gauss-relat}
S_r(t,\mathbf{r})=\frac{1}{16\pi^2 R_x R_y R_z \tau} \:
\exp\Big[ -\frac{t ^2}{4\tau^2} -\frac{x^2}{4R_x^2}
-\frac{y^2}{4R_y^2}-\frac{z^2}{4R_z^2}\Big].
\ee
It is worth noting that the particle masses, that are present in the definition (\ref{source-relat}), disappear completely in the formula (\ref{gauss-relat}). This is the feature of the Gaussian parameterization (\ref{gauss}). 

If a pair of particles moves relative to the source along the axis $x$ with the velocity ${\bf v} = (v,0,0)$, the effective relative source function (\ref{effective-r-source}), which will be discussed later on, is
\be
\label{effective-r-source-Gauss-vx}
S_r^{\rm eff}({\bf r}) = \frac{1}{8\pi^{3/2}\sqrt{(R_x^2+v^2\tau^2)}R_yR_z}
\exp \left[-\frac{1}{4}
\left(\frac{x^2}{R_x^2+v^2\tau^2}
+\frac{y^2}{R_y^2}+\frac{z^2}{R_z^2}\right)\right] .
\ee

If the velocity ${\bf v}$ is perpendicular to the beam axis along the axis $z$, as in my further considerations, the Cartesian coordinates $x$, $y$, $z$ is a special case of the Bertsch-Pratt coordinates system {\it out}, {\it side}, {\it long} \cite{Bertsch:1988db,Pratt:1986cc} often used in femtoscopy. The direction {\it long} is chosen along the beam axis ($z$), the {\it out} is parallel to the component of the pair velocity ${\bf v}$ which is transverse to the beam. The last direction {\it side} is along the vector product of the {\it out} and {\it long} versors. Further on, the direction along the motion of the pair is called the `out' direction. 

The Gaussian parameterization of the source function is convenient to use and, as I show in Sec.~\ref{sec-source-function}, allows one to readily include relativistic effects. However, the Gaussian character of the source is not a law of nature, but merely a useful approximation. In this context, it is worth mentioning the technique called {\it imaging}, proposed by Brown and Danielewicz \cite{Brown:1997ku,Brown:1997sn}. For systems such as pairs of pions or protons, whose two-particle wave functions are known, one can invert the Koonin formula (\ref{Koonin-r-3D}) and obtain the effective source function (\ref{effective-r-source}) from the experimentally measured correlation function. Applying the imaging technique to experimental data has shown that the source functions are indeed approximately Gaussian, but with additional non-Gaussian tails \cite{Brown:2000aj,Verde:2001md,PHENIX:2006nml,PHENIX:2007grx,NA49:2008ofa}. As I will explain in Sec.~\ref{sec-source-function}, the source functions obtained using imaging technique do not allow the analysis of relativistic effects, which are of primary interest here.

As already mentioned in the introduction, for non-interacting particles the Koonin formula (\ref{Koonin}) can be written in a relativistically covariant form with time and space variables treated on equal terms, see {\it e.g.} \cite{Wiedemann:1999qn}. This formula can be used to compute the correlation function of free identical bosons or fermions which differs from unity due to the effect of quantum statistics. It is worth noting that, as discussed in \cite{Maj:2009ue}, these correlation functions are exactly equal to the analogous functions provided by the Koonin formula (\ref{Koonin}). The equality is not obvious as the time enters the Koonin formula and its relativistic generalization in a different way. So, it confirms the validity of the simple geometric arguments that allow for the inclusion of non-simultaneous particle emission in the Koonin formula.

\section{Relativistically covariant source function}
\label{sec-source-function}

To apply the Koonin formula (\ref{Koonin}), (\ref{Koonin-r-4D}) or (\ref{Koonin-r-3D}) to pairs of correlated particles which move relativistically with respect to a source, one has to transform the source function, which is written down as $S(x)$ with the four-position $x \equiv (t,{\bf r})$, from the source rest frame to the pair CM frame. The source function $S(x)$ is the probability density of the emission of a particle from a given space-time point $x$, and thus the emission probability $S(x)\,d^4x$ should be the same for all observers, it should be Lorentz invariant. 

When the reference frame ${\cal O}'$ moves along the axis $x$ of the reference frame ${\cal O}$ with the velocity $v$ and initially, that is at $t=0$, the two frames coincide, the position four-vector expressed in the Minkowski coordinates transforms as
\be
\label{Lorentz-boost}
x'^\mu = L^\mu_{\;\;\nu} x^\nu =
\left[ \begin{array}{c c c c}
    \gamma   &  - \gamma v  & 0 & 0 \\
 - \gamma v &  \gamma       & 0 & 0 \\ 
             0   &          0         & 1 & 0 \\
             0   &          0         & 0 & 1 \\
\end{array} \right]
\left[ \begin{array}{c}
t \\ x \\ y \\ z \\ 
\end{array} \right]
= \left[ \begin{array}{c}
\gamma(t - v x) \\ \gamma(x - v t) \\ y \\ z \\ 
\end{array} \right] .
\ee
where $\gamma \equiv (1-v^2)^{-1/2}$. The indices $\mu$ and $\nu$ numerate the rows and columns, respectively. The matrix $L$ is symmetric that is $L_\mu^{\;\;\nu}=L_\nu^{\;\;\mu}$ and it obeys $(L^{-1})^\mu_{\;\;\nu} = L_\mu^{\;\;\nu}$ and $L_\mu^{\;\;\nu} L^\mu_{\;\;\rho} = g^\nu_{\;\;\rho}$. 

Since the four-volume element is Lorentz invariant $d^4x' = d^4x$, the emission probability is the same for all observers if the source function transforms as a scalar field that is
\be
\label{scalar-transform}
S(x) \longrightarrow S'(x') = S(L^{-1}x') . 
\ee

The explicitly covariant form of the Gaussian parameterization of the source function (\ref{gauss}) can be written as \cite{Heinz:1996qu}
\be
\label{source-cov}
S(x)=\frac{\sqrt{{\rm det}\Lambda}}{4\pi^2} \;
{\rm exp} \Big[-\frac{1}{2}x_\mu \Lambda^{\mu\nu}x_\nu \Big],
\ee
where $\Lambda^{\mu\nu}$ is assumed to be a Lorentz tensor characterizing the source which in the source rest frame is
\be
\label{source-matrix}
\Lambda^{\mu\nu}=\left[\begin{array}{cccc}
\frac{1}{\tau^2} & 0 & 0 & 0  \\
0 & \frac{1}{R_{x}^2} & 0 & 0  \\
0 & 0 & \frac{1}{R_{y}^2} & 0  \\
0 & 0 & 0 & \frac{1}{R_{z}^2}  \\
\end{array}\right].
\ee
The source function as written in Eq.~(\ref{source-cov}) obeys the normalization condition (\ref{norma}) not only for the diagonal matrix $\Lambda$ but for non-diagonal as well. 

The source function (\ref{source-cov}) obviously transforms according to the rule (\ref{scalar-transform}) that is 
\be
S(L^{-1}x') = \frac{\sqrt{{\rm det} \Lambda'}}{4 \pi^2}
\exp \Big[-\frac{1}{2} {x'}_\rho {\Lambda'}^{\rho\sigma} {x'}_\sigma \Big] 
= S'(x') \,,
\ee
where 
\be
\label{transform-Lambda}
{\Lambda'}^{\rho\sigma} = L^\rho_{\;\;\mu}\Lambda^{\mu\nu} (L^{-1})_\nu^{\;\;\sigma} 
= L^\rho_{\;\;\mu} L^\sigma_{\;\;\nu}  \Lambda^{\mu\nu} 
\ee
and 
${\rm det} \Lambda' = {\rm det}L \: {\rm det} \Lambda \: {\rm det}L^{-1} = {\rm det} \Lambda$. Once the source function (\ref{source-cov}) transforms as a scalar field, it is known how to write it in any reference frame. 

Let me write explicitly the relative source function (\ref{gauss-relat}) in the CM frame of the particle pair, which is assumed to move with the velocity ${\bf v} = (v,0,0)$ in the source rest frame. The quantities are labeled with the index $*$ in the CM frame. Using the transformation law (\ref{transform-Lambda}) and the Lorentz boost matrix from Eq.~(\ref{Lorentz-boost}), one finds
\ba
\label{source-matrix-CM}
\Lambda^{\mu\nu}_*=\left[\begin{array}{cccc}
\gamma^2(\frac{1}{\tau^2}+\frac{v^2}{R_{x}^2}) & 
- \gamma^2v(\frac{1}{\tau^2}+\frac{1}{R_{x}^2}) & 0 & 0  \\
-\gamma^2v(\frac{1}{\tau^2}+\frac{1}{R_{x}^2}) & 
\gamma^2(\frac{v^2}{\tau^2}+\frac{1}{R_{x}^2}) & 0 & 0  \\
0 & 0 & \frac{1}{R_{y}^2} & 0  \\
0 & 0 & 0 & \frac{1}{R_{z}^2}  \\
\end{array}\right] .
\ea
The source function is
\be
\label{r-source-CM}
S_r(x^*) = \frac{1}{16\pi^2 \tau R_x R_y R_z}
\exp \Big[-\frac{1}{4} x^*_\mu \Lambda_*^{\mu\nu}x^*_\nu \Big] . 
\ee
Finally, one writes down the `effective relative' source function defined by Eq.~(\ref{effective-r-source}) which is a relativistic analogue of the formula (\ref{effective-r-source-Gauss-vx}) and equals 
\be
\label{effective-r-source-CM}
S_r^{\rm eff}({\bf r}_*) \equiv \int dt_* \, S_r(t_*,{\bf r}_*) =
\frac{1}{8\pi^{3/2}\sqrt{\gamma^2(R_x^2+v^2\tau^2)}R_yR_z}
\exp \left[-\frac{1}{4}
\left(\frac{x_*^2}{\gamma^2(R_x^2+v^2\tau^2)}
+\frac{y_*^2}{R_y^2}+\frac{z_*^2}{R_z^2}\right)\right] ,
\ee
where ${\bf v}^* =0$ and ${\bf v} = (v,0,0)$. As one observes, the effective source radius along the direction of the velocity is elongated by the factor $\gamma$, not contracted as one can naively expect. It should be remembered that the Lorentz contraction of the length of a moving object occurs if both ends of the object are measured simultaneously in the reference frame in which the object is moving. Changing the measurement method would lead to a different result, and this is the case here.

The formula (\ref{effective-r-source-CM}) is the source function to be substituted into Eq.~(\ref{Koonin-r-3D}) to get the correlation function in the CM frame of a particle pair. It will be used further on. 

Let me return for a moment to the imaging technique mentioned in the previous section. The technique allows the extraction of effective source functions from experimental correlation functions. Since correlation functions are measured in the CM frame of correlated particles, the source functions obtained by inverting the Koonin formula are expressed in this reference frame. They cannot be Lorentz transformed to the source rest frame because the emission time is integrated over in the definition of the effective source function (\ref{effective-r-source}).

\section{Magnitude of relativistic effects}
\label{sec-magnitude}

As already explained, the main relativistic effect in femtoscopy is due to Lorentz transformation of a source function. The source radius in the direction of particle motion, which in the source rest frame is $\sqrt{R_x^2 + v^2\tau^2}$, is enlarged by the Lorentz factor $\gamma$. In this section I discuss how much it influences the correlation function. 

If the correlation function is measured for particle pairs within a narrow range of total momentum ${\bf P}$, relativistic effects only affect the interpretation of the obtained source radius. The source radius in the direction of particle motion, which is inferred from the correlation function measured in the CM frame of the particle pair, is $\gamma \sqrt{R_x^2 + v^2\tau^2}$. A situation gets complicated if the measurement is performed in a broader range of total momentum ${\bf P}$. Then, one deals with a correlation function averaged over a domain of ${\bf P}$.

I start the discussion on a magnitude of relativistic effects with $\Lambda$-$p$ correlations which are caused by short-range strong interaction and thus the correlation function is sensitive to the source radius. However, a measurement of the $\Lambda$-$p$ correlation function is rather difficult as it requires a reconstruction of  $\Lambda$ from the decay products $p$ and $\pi^-$. So, $p$-$p$ correlations, which are easier to measure, are also discussed. However, the $p$-$p$ correlations are strongly influenced by the long range Coulomb repulsion which makes the correlation function less sensitive to the source radius, in particular when the source radius exceeds 2-3 fm as in heavy-ion collisions. 

An observation of relativistic increase of the source radius in the out direction is complicated by the fact that the source radius changes with ${\bf P}$ not only due to the Lorentz transformation. It is well established that the source radius inferred from the femtoscopic correlation function decreases with increasing transverse momentum of particle pairs, see {\it 
e.g.} Fig.~8 of \cite{ALICE:2015hvw}. This behavior, which is known as the {\it homogeneity length effect}, is a consequence of the hydrodynamic radial expansion of matter produced in high-energy collisions \cite{Akkelin:1995gh}. A rather detailed analysis of the phenomenon in a hydrodynamical model is given in \cite{Kisiel:2014upa}. Taking into account the relativistic increase of the source radius with particle momentum, the effect becomes stronger than currently assumed if the correlation function is measured in the CM frame of the pair of correlated particles. However, to disentangle the effect of Lorentz transformation from the genuine change of the source radius with the particle momentum, the correlation function must be measured in small intervals of the pair total momentum ${\bf P}$ or the pair velocity ${\bf v}$.

At the end of this section, $\pi^\pm$-$\pi^\pm$ and $K^\pm$-$K^\pm$ correlations are considered. The experimental data are traditionally analyzed and presented in a specific way. Using the fact that the strong interaction of pions and kaons can be neglected and treating the Coulomb repulsion as a correction, the final correlation functions are supposed to represent free particles with non-trivial correlations caused solely by quantum statistics. I explain how relativistic effects are handled and argue that the method works well for pions but is less accurate for kaons. 

The correlations functions discussed here are, of course, computed in the CM frame. However, to simplify the notation asterisks to denote quantities in the CM frame are no longer used, except in a few cases where I wish to emphasize this fact.

\subsection{$\Lambda$-$p$ correlations}

I discuss here specific measurements of $\Lambda$-$p$ correlations performed by STAR Collaboration in Au-Au collisions at $\sqrt{s_{NN}}=200$~GeV \cite{STAR:2005rpl} and by ALICE Collaboration in $p$-$p$ collisions at $\sqrt{s_{NN}}=7$~TeV \cite{ALICE:2018ysd}. 

In case of STAR measurement the protons were from the transverse-momentum interval $p_T \in [0.4,1.1]\,$GeV and rapidity $y \in [-0.5,0.5]$ in the CM frame of the colliding nuclei which is assumed to be the rest frame of the source. For $y=0$ the Lorentz factor varies from $\gamma = 1.09$ for $p_T = 0.4\,$GeV to $\gamma = 1.54$ for $p_T = 1.1\,$GeV. Therefore, the source radius in the direction of particle pair motion increases by over 40\% when $p_T$ grows from 0.4 to 1.1 GeV. However, since the particle yield exponentially drops with $p_T$, the correlation function is mostly determined by the pairs with the lowest $p_T$ and the spread of source radii due to the Lorentz transformation is effectively much smaller than 40\%. 

The measurement by ALICE Collaboration was performed with protons from the interval $p_T \in [0.5,4.05]\,$GeV and pseudorapidity $\eta \in [-0.8,0.8]$. Assuming as previously $\eta = 0$,  the Lorentz factor varies from $\gamma = 1.13$ for $p_T = 0.5\,$GeV to $\gamma = 4.4$ for $p_T = 4.05\,$GeV. Therefore, the source radius in the direction of particle pair motion increases by factor of 4 when $p_T$ grows from 0.5 to 4.05 GeV. 

The correlation functions obtained by STAR and ALICE Collaborations were compared with theoretical predictions both assuming that the source function was isotropic in the CM frame of $\Lambda$-$p$ pairs. This means $\gamma \sqrt{R_x^2 + v^2\tau^2} =  R_y = R_z \equiv R_s$. The source radius was inferred to be $R_s = 3.0 \pm 0.7\,$fm for Au-Au collisions and $R_s = 1.13 \pm 0.02\,$fm for $p$-$p$ collisions. However, the source radius in the out direction grows with the total momentum of the pair due to the Lorentz transformation. The question is how it influences the correlation function.

The correlation function is computed using the Lednick\'y-Luboshitz method \cite{Lednicky:1981su,Gmitro:1986ay,Lednicky:2005tb}. It assumes that the wave function of correlated particles, which enters the formula (\ref{Koonin-r-3D}), is in the scattering asymptotic form. When the interparticle interaction is short range the wave function is
\be
\label{wave-fun-asym-scatt}
\varphi_{\bf q}({\bf r})=e^{i{\bf q}{\bf r}}+f(q)\frac{e^{iqr}}{r},
\ee
where $f(q)$ is the $s$-wave scattering amplitude given as
\be 
\label{ampli}
f (q)  = \frac{- a}{1 - \frac{1}{2} d a q^2 + i q a} , 
\ee
where $a$ and $d$ are the scattering length and effective range. The modulus squared of the wave function (\ref{wave-fun-asym-scatt}) together with the source function (\ref{effective-r-source-CM}) are substituted in Eq.~(\ref{Koonin-r-3D}) and integrated over the position ${\bf r}$. For an isotropic source function, the integral can be performed analytically to some extent whereas an anisotropic source function requires a fully numerical three-dimensional integration. 

Since the both proton and $\Lambda$ are spin 1/2 particles, the pair can be in a singlet and triplet state. Assuming that the particles are unpolarized, the spin averaged correlation function is
\be 
\label{spin-ave} 
C(\mathbf{q},\mathbf{P}) = \frac{1}{4}\, C^s(\mathbf{q},\mathbf{P}) 
+ \frac{3}{4} \, C^t(\mathbf{q},\mathbf{P}) ,
\ee
where $C^{s,t}(\mathbf{q},\mathbf{P})$ is the correlation function of the pair in the singlet and triplet state, respectively, and the weights 1/4 and 3/4 reflect the numbers of spin states in the two channels. 

\begin{figure}[t]
\centering
\includegraphics[width=11cm]{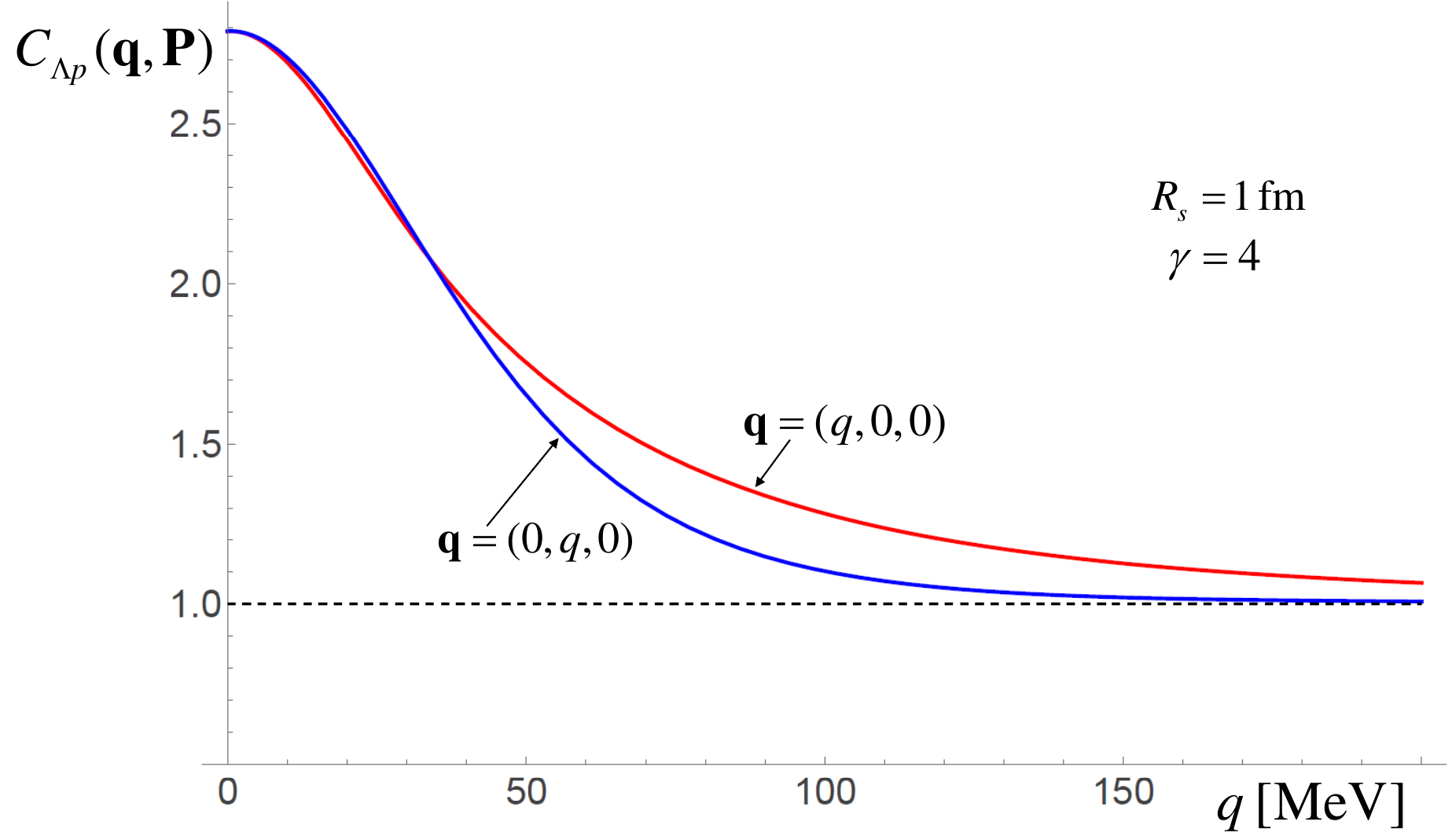}
\vspace{-3mm}
\caption{The $\Lambda$-$p$ correlation function as a function of ${\bf q}=(q,0,0)$ (red) and of ${\bf q}=(0,q,0)$ (blue)  computed for $R_s = 1\,$fm and $\gamma = 4$.}
\label{fig-lambda-p}
\end{figure}

Using the singlet and triplet scattering length and effective range 
\be
a^s = -2.91~{\rm fm},~~~ d^s  = 2.78~{\rm fm},
~~~~~~~~~~~~~~~~
a^t = -1.54~{\rm fm},~~~ d^t  = 2.72~{\rm fm},
\ee
computed in \cite{Haidenbauer:2013oca} within the chiral effective theory, the spin averaged correlation function has been computed, assuming that in the source rest frame the source is isotropic with $R_x =R_y = R_z \equiv R_s$ and the emission time $\tau$ has been set to zero, so it is assumed $\tau \ll R_s$. As mentioned above, the $\Lambda$-$p$ pairs with protons of the transverse momentum as high as $p_T = 4.05\,$GeV contributed to the $\Lambda$-$p$ correlations measured by the ALICE Collaboration \cite{ALICE:2018ysd}. So, the correlation function is computed for $R_s = 1\,$fm and $\gamma=4$ and the result is shown in Fig.~\ref{fig-lambda-p}. Since the source is anisotropic in the CM frame of the $\Lambda$-$p$ pairs, the correlation function depends on the orientation of the momentum ${\bf q}$. Fig.~\ref{fig-lambda-p} shows the correlation functions for  ${\bf q}=(q,0,0)$ and ${\bf q}=(0,q,0)$. In the domain of small $q$ ($q < 30\,$MeV) both correlation functions coincide but when $q$ grows the correlation function of ${\bf q}=(0,q,0)$ tends to unity faster than that of ${\bf q}=(q,0,0)$. Taking into account that the source radius in the direction of particle motion is four times bigger than that in the perpendicular direction the difference between the correlation functions of ${\bf q}=(0,q,0)$ and ${\bf q}=(q,0,0)$ is not very sizable. The difference is even smaller for smaller $\gamma$ and it obviously vanishes as $\gamma \to 1$.

The correlation functions from Fig.~\ref{fig-lambda-p} show that the relativistic effect cannot be easily observed. First of all, the correlation function should be measured in narrow intervals of the pair momentum ${\bf P}$. Then, one could study the dependence of the correlation function on $\gamma$. Additionally, the correlation function should measured as a function the vector ${\bf q}$ not only $|{\bf q}|$. This is certainly a difficult task and interpretation of the results of such a study will be complicated by the already mentioned fact that the source radius in the out direction changes with the pair transverse momentum not only due to the relativistic elongation but also due to the radial expansion of the source. 

In spite of all these difficulties, it should be remembered that the relativistic effects are present even when one is unable to single them out. The problem is particularly severe when the correlations function is measured in a broad range of ${\bf P}$ and as a function $|{\bf q}|$ not ${\bf q}$, as it is done in the studies \cite{STAR:2005rpl} and \cite{ALICE:2018ysd}. Then, the resulting correlation function is averaged over various $\gamma$ factors and orientations of the vector ${\bf q}$. In case of the  $\Lambda$-$p$ correlations it is difficult to go beyond these limitations. So, I consider the $p$-$p$ correlations which are easier to measure. 

\subsection{$p$-$p$ correlations}

The  $p$-$p$ correlation function is easier to measure but it noticeably more difficult to compute, as the inter-particle strong interaction and Coulomb repulsion must be both taken into account. 

The Lednick\'y-Luboshitz method \cite{Lednicky:1981su,Gmitro:1986ay,Lednicky:2005tb} is again applied. However, the scattering asymptotic function is no longer of the from (\ref{wave-fun-asym-scatt}), as the long range electromagnetic interaction modifies both the incoming plane wave and outgoing spherical wave even far away from the scattering center. Since the particle source is always much smaller than the Bohr radius, which characterizes the length scale of the electromagnetic interaction, the wave function of interest has to be of the asymptotic from the point of view of the short range strong interaction but it must describe exactly the electromagnetic interaction. In case of non-identical particles the wave function, which is given by Eq.~(12) in \cite{Gmitro:1986ay} and Eq.~(89) in \cite{Lednicky:2005tb}, equals 
\be 
\label{wave-function-coulomb-strong-G} 
\varphi_{\bf q}({\bf r})
= e^{i\delta_0^C} \, \sqrt{G(q)} \,  e^{i\bf qr} \,
_1F_1\Big(-{i \over a_B q}, 1, i q \eta \Big)  
+  e^{i\delta_0^C} \tilde{f}(q) \, \frac{1}{r}
 \Big[G_0 \Big(\frac{1}{a_B q},qr\Big) + i F_0 \Big(\frac{1}{a_B q},qr\Big) \Big] , 
\ee
where $q \equiv |{\bf q}|$, $\eta \equiv r - z = r(1 - \cos\theta)$ and $a_B \equiv (\mu \alpha)^{-1}$ is the Bohr radius with $\mu$ being the reduced mass of the two particles and $\alpha = 1/137$ is the fine structure constant; $_1F_1(a,b,z)$ denotes the hypergeometric confluent function and $G(q)$ is the Gamow factor 
\be 
\label{Gamow}
G(q) \equiv |\varphi_{\bf q}^C({\bf r}=0)|^2 
= {2 \pi \over a_B q} \,{1 \over {\rm exp}\big({2 \pi \over a_B q}\big) - 1} 
\ee
with $\varphi_{\bf q}^C({\bf r})$ being the Coulomb wave function (\ref{coulomb-wave-2}) and
\be
\delta_0^C \equiv {\rm arg}\Big[\Gamma \Big(1 +{i \over a_B q} \Big)\Big] 
\ee
is the $s$-wave Coulomb phase-shift; $\tilde{f}$ is the Coulomb modified scattering amplitude 
\be
\label{f-0}
\tilde{f}(q) = \frac{-a \, G(q)}{1 - \frac{1}{2} d a q^2 + \frac{2 a}{a_B} \, h(a_B q) + i aq \, G(q)} ,
\ee
where the function $h(x)$ is defined as 
\be
\label{h-def}
h(x) = \frac{1}{x^2} \sum_{n=1}^\infty \frac{1}{n (n^2 + x^{-2})} - C + \ln x 
\ee
with $C \approx 0.577$ being Euler's constant; $F_0$ and $G_0$ are the $l=0$ Coulomb functions which are, respectively, regular and irregular at $r=0$. The first term of the formula (\ref{wave-function-coulomb-strong-G}) is the exact scattering Coulomb wave function while the second term represents the $l=0$ spherical wave which results from the strong interaction combined with the Coulomb repulsion. 

The wave function (\ref{wave-function-coulomb-strong-G}) was derived in \cite{Lednicky:1981su,Gmitro:1986ay,Lednicky:2005tb} in the context relativistic Bethe-Salpeter equation which is, in my opinion, somewhat misleading, as the approximations which lead to the final formula of the correlation function hold when the two particles of interest are non-relativistic. The description of the derivation of the wave function (\ref{wave-function-coulomb-strong-G}) given in \cite{Lednicky:1981su,Gmitro:1986ay,Lednicky:2005tb} is also very brief. Therefore, I have rederived the formula (\ref{wave-function-coulomb-strong-G}) and present a fairly detailed description of the derivation in Appendix.

\begin{figure}[t]
\centering
\includegraphics[width=11cm]{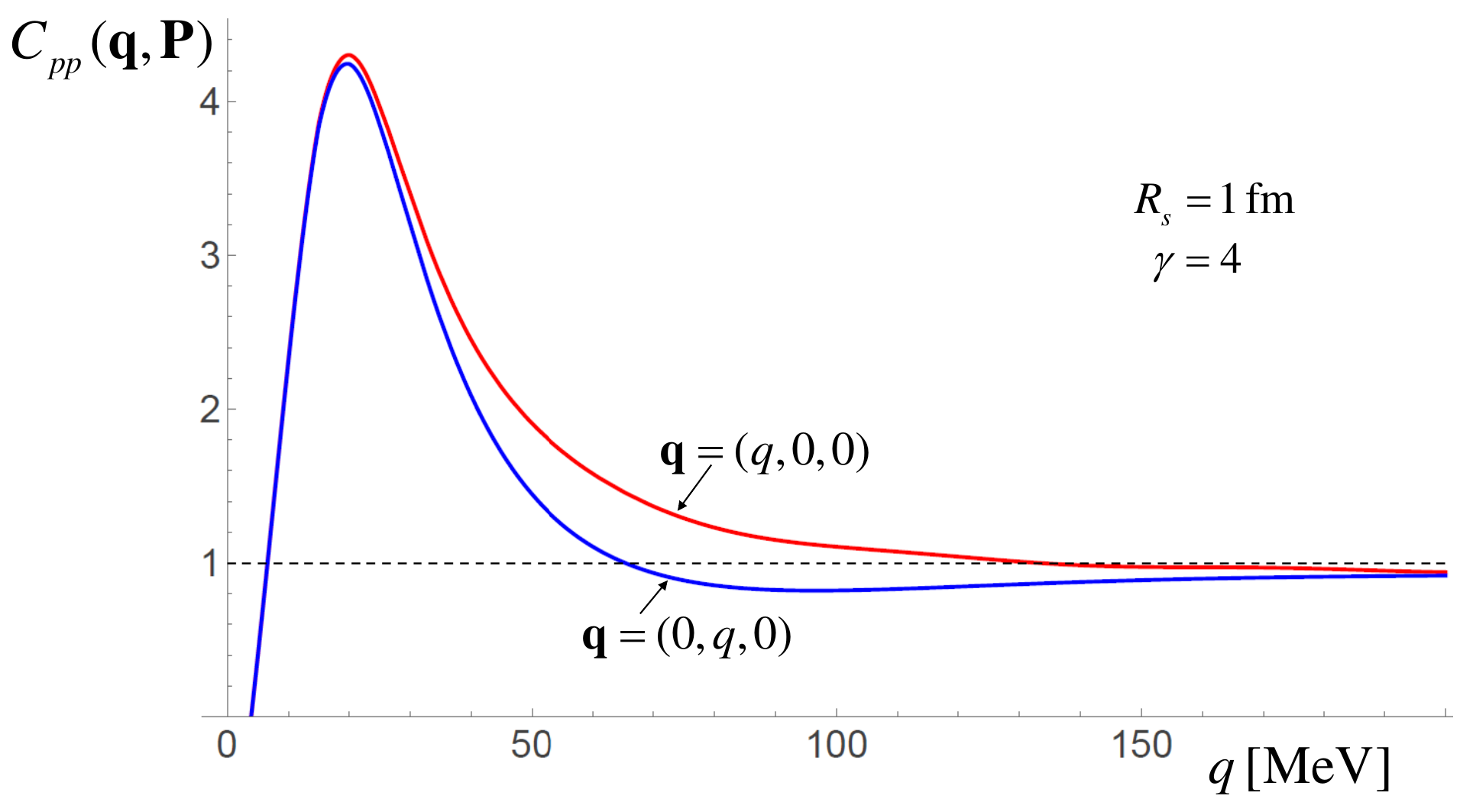}
\vspace{-3mm}
\caption{The $p$-$p$ correlation function as a function of ${\bf q}=(q,0,0)$ (red) and of ${\bf q}=(0,q,0)$ (blue)  computed for $R_s = 1\,$fm and $\gamma = 4$.}
\label{fig-p-p}
\end{figure}

Pairs of protons, as $\Lambda$-$p$ pairs, can be in the spin singlet and triplet sate. Since one deals with pairs of identical fermions, the complete wave function must be antisymmetric when the two particles are interchanged that is under the replacement ${\bf r} \to -{\bf r}$. The spin triplet ($S=1$) wave function is symmetric, as the particles are in the same spin state, while the spin singlet ($S=0$) wave function is antisymmetric. Consequently, the spatial wave  function of the triplet is antisymmetric and that of the singlet symmetric. The triplet and singlet wave function to be substituted in the Koonin formula (\ref{Koonin-r-3D}) are
\ba
\label{wave-t}
\varphi_{\bf q}^t({\bf r}) = \frac{1}{\sqrt{2}}\Big(\varphi_{\bf q}({\bf r}) - \varphi_{\bf q}(-{\bf r})\Big),
\\
\label{wave-s}
\varphi_{\bf q}^s({\bf r}) = \frac{1}{\sqrt{2}}\Big(\varphi_{\bf q}({\bf r}) + \varphi_{\bf q}(-{\bf r})\Big),
\ea
where the wave function $\varphi_{\bf q}({\bf r})$ is given by Eq.~(\ref{wave-function-coulomb-strong-G}). The second term in Eq.~(\ref{wave-function-coulomb-strong-G}) which describes the $l=0$ outgoing spherical wave is symmetric under the replacement ${\bf r} \to -{\bf r}$. Therefore, it does not contribute to $\varphi_{\bf q}^t({\bf r})$ that is the triplet wave function is determined solely by the Coulomb interaction. So, to compute the $p$-$p$ one needs only singlet scattering length and effective range which according to the textbook \cite{Landau-Lifshitz-1988} are
\be
a^s = - 7.8~{\rm fm},~~~ d^s  = 2.8~{\rm fm}.
\ee
The Bohr radius of a proton pair is $a_B = 57.6\,$fm. 

One gets the triplet and singlet correlation function substituting the modulus square of the wave functions (\ref{wave-t}) and (\ref{wave-s}) into Eq.~(\ref{Koonin-r-3D}), and using the formula (\ref{spin-ave}) the spin average correlation function is obtained. As in case of $\Lambda$-$p$ correlations, it is assumed that the source in its rest frame is isotropic with $R_s = 1\,$fm and the gamma factor of the Lorentz transformation to the pair CM frame is $\gamma = 4$. If the source function is isotropic in the CM frame of the particle pair, the integral can be performed analytically to some extent. In case of an anisotropic source, one has to rely on a fully numerical three-dimensional integration. 

Fig.~\ref{fig-p-p} shows the correlation functions for ${\bf q}=(q,0,0)$ and ${\bf q}=(0,q,0)$. For small $q$ ($q < 30\,$MeV) both correlation functions are close to each other but when $q$ grows the correlation function of ${\bf q}=(0,q,0)$ is significantly bigger than that of ${\bf q}=(q,0,0)$ and then they both tend to unity from below. As in case of $\Lambda$-$p$ correlations, the difference between the correlation functions is not very sizable.

\subsection{$\pi$-$\pi$ and $K$-$K$ correlations}

Relativistic effects are most important for pions, as the lightest hadrons. In case of the correlation measurement of like-sign pions \cite{ALICE:2015tra}, the pions of transverse momenta up to 1 GeV were included in the analysis. So, the gamma factor of the pion transverse motion was extended up to $\gamma = 7.2$ which shows that the relativistic effects are indeed sizable. However, these effects show up differently than in $p$-$p$ and $\Lambda$-$p$ correlations, as those of $\pi$-$\pi$ and $K$-$K$ are measured, analyzed and presented differently.

The  method to study the $\pi^\pm$-$\pi^\pm$ correlations exploits the fact the strong interaction can be neglected while the Coulomb repulsion is treated as a correction subtracted from experimental data. The final correlation function $C_{\rm free}(\mathbf{q},\mathbf{P})$ is expected to represent correlations of free pions. This function, which is easily computed substituting the symmetrized plane wave and source function (\ref{effective-r-source-Gauss-vx}) into Eq.~(\ref{Koonin-r-3D}), equals
\be 
\label{corr-fun-free} 
C_{\rm free}(\mathbf{q},\mathbf{P}) 
= 1 + \exp \big[-4 \big( \gamma^2(v^2\tau^2 + R_x^2)q_x^2 
+ R_y^2 q_y^2+ R_z^2 q_z^2\big)\big]\; .
\ee 
It can be also written in an explicitly Lorentz covariant form \cite{Heinz:1996qu} as
\be 
\label{corr-fun-free-cov} 
C_{\rm free}(Q) 
= 1 + \exp [-4Q_\mu(\Lambda^{-1})^{\mu\nu} Q_\nu ] ,
\ee 
where $Q = ({\bf q}\cdot {\bf v}, {\bf q})$ is the four-vector and $\Lambda^{-1}$ is the tensor inverse to $\Lambda$ 
given by Eq.~(\ref{source-matrix}) in the source rest frame. As mentioned at the end of Sec.~\ref{sec-koonin}, the correlation function (\ref{corr-fun-free}) can be also obtained from the relativistic generalization of the Koonin formula where time and space coordinates are treated on equal terms. 

The simplest version of the method to study the $\pi^\pm$-$\pi^\pm$ assumed that the correlation function was simply divided by the Gamow factor (\ref{Gamow}), see {\it e.g.} \cite{WA80:1992hei}. Then, the Bose-Einstein enhancement at the smallest relative momenta of pions, which was reduced by the Coulomb repulsion, was clearly seen. As one remembers, the Gamow factor is the modulus squared of the Coulomb wave function at ${\bf r}=0$ given by Eq.~(\ref{Gamow}). The method is justified by the fact that the Bohr radius of two pion system $a_B = 386\,$fm, which is the characteristic length scale of electromagnetic interaction, is much bigger that the size of the pion source in high-energy collisions. 

Since the Gamow factor overestimates the Coulomb effect of particles emitted from a finite-size source, the simple division of the correlation function by the Gamow factor has been replaced by the Bowler-Sinyukov method \cite{Bowler:1991vx,Sinyukov:1998fc}, according to which the measured correlation function is expressed as follows
\be
\label{B-S-method}
C(\mathbf{q},\mathbf{P}) =  K(q_*) \, C_{\rm free}(\mathbf{q},\mathbf{P}) ,
\ee
where $C_{\rm free}(\mathbf{q},\mathbf{P})$ is the correlation function of free pions (\ref{corr-fun-free}) and $K(q_*)$ is the correction factor  whose role is to take into account the Coulomb effect. $K(q_*)$ is actually the correlation function of two distinguishable particles of pion masses computed in the CM frame of pion pair with the Coulomb wave function (the first term in the formula (\ref{wave-function-coulomb-strong-G})) and the istropic source function of radius $R_s$. The radius is chosen to be close to the radii which enter the free correlation function. The formula (\ref{B-S-method}) usually includes the parameter $\lambda$ which is introduced to eliminate some experimental problems like particle missidentification or an effect of coherence of a pion field. As these issues are irrelevant to my considerations, I set $\lambda = 1$. 

The Bowler-Sinyukov method assumes that the Coulomb contribution to a correlation function is isotropic even when the source is anisotropic and the correlation function depends on the orientation of ${\bf q}$. The assumption was tested and for pions is well satisfied \cite{Maj:2009ue}. However, the formula (\ref{B-S-method}) is less accurate for kaons, in particular when sources are as big as in central collisions of heavy-ions \cite{Maj:2009ue}. This happens because the Bohr radius of $K^\pm$-$K^\pm$ pair $a_B = 110\,$fm is 3.5 times smaller than that of $\pi^\pm$-$\pi^\pm$ pair and the electromagnetic forces become sensitive to a shape of the source. 

Since relativistic effects are of main interest here, the Lorentz structure of the formula (\ref{B-S-method}) needs to be discussed. Although the correction factor $K(q_*)$ is computed in the CM frame of a pion pair, the formula can be used in any reference frame. The factor $K(q_*)$, as a correlation function, is a Lorentz scalar function. Actually, the function does not need to be transformed because the argument $q_*$ is Lorentz invariant. The point is that $q_*  \equiv |{\bf q}_*| = \sqrt{ - Q^\mu Q_\mu}$ where $Q^\mu \equiv \frac{1}{2}(p_1^\mu - p_2^\mu)$ and $p_1^\mu, \, p_2^\mu$ are four-momenta of the two pions. Since $Q^0 = {\bf q} \cdot {\bf v}$, it vanishes in the CM frame of the pair and thus $q_*  = \sqrt{{\bf q}_*^2} = \sqrt{ - Q^\mu Q_\mu}$ which is Lorentz invariant.

The formula (\ref{B-S-method}) is approximate but it is Lorentz covariant, it holds in any reference frame. The formula is applied to three-dimensional measurements of the correlation functions which show how these functions depend of the orientation of the vector ${\bf q}$. Such measurements, see {\it e.g.} \cite{ALICE:2015tra}, are usually performed using the Bertsch-Pratt coordinates in the longitudinally comoving reference frame where rapidity of the particle pair vanishes. 

Although the Bowler-Sinyukov method is certainly useful, one should remember that it was proposed long ago when correlation functions were measured with lower accuracy. Now, the measurements are much more precise and it is hard to justify why the final result of the analysis are the correlation function with the subtracted electromagnetic contribution. It is rather unfortunate that experimental data on $\pi$-$\pi$ and $K$-$K$ are presented differently than those on, say, $p$-$p$ and $\Lambda$-$p$. This is particularly striking when the two methods are used in the same study, as in \cite{ALICE:2015hvw}. The most complete information is contained in the correlation functions measured as a function of both ${\bf q}$ and ${\bf v}$, or equivalently as a function of four-vector $Q$.  In particular, they allow one to separate the effects of relativistic elongation of the source radius from its dynamical change. Hopefully, such measurements will be available not in a remote future.

\section{Deuteron formation}
\label{sec-deuteron}

As already mentioned in the introduction, production of deuterons in high-energy collisions is a two-step process in the coalescence model \cite{Butler:1963pp,Schwarzschild:1963zz}. At first nucleons are produced and later on a formation of deuterons proceeds due to final-state interactions of nucleons which are close to each other in momentum and coordinate spaces. The fact that the energy scale of the first step, which is of order of nucleon mass, is much higher than that of the second one, which is a deuteron binding energy, is important for two reasons. First of all, the two steps of the process can be distinguished because the temporal scales are roughly inverse of the energy scales. Consequently, the production of nucleons, which occurs at first, is a fast process while the formation of deuterons is a slow process which occurs subsequently. Secondly, a probability to produce a  deuteron can be factorized into the probability to produce a neutron-proton pair and the probability that nucleons fuse into a deuteron. 

\subsection{Non-relativistic formulation}

Assuming that the neutron and proton observed in a given collision are produced independently of each other and that the neutron multiplicity is equal to the proton multiplicity, the deuteron yield is given as 
\be
\label{A-def}
\frac{dN_d}{d^3p_d} = A \, \Big(\frac{dN_p}{d^3p_p} \Big)^2,
\ee
where $A$ is the deuteron formation rate and ${\bf p}_d = 2{\bf p}_p$. One ignores here a small effect of the deuteron binding energy and thus, the equality  ${\bf p}_d = 2{\bf p}_p$ is only approximate. However, it does not influence the yield of neutron-proton pairs which changes at a much bigger energy scale than the deuteron binding energy. 

A third body is not needed to convert a neutron-proton pair into a deuteron. As observed long ago \cite{Mrowczynski:1987oid}, nucleons, which are emitted from a fireball, are {\em not} on the mass shell due to the finite space-time size of a fireball. The space-time localization of a nucleon within the fireball washes out its four-momentum due to the uncertainty principle. Using a more formal language of scattering theory, the nucleons are not in an asymptotic state in the remote past or remote future which indeed requires the mass-shell condition. Instead the nucleons are in an intermediate scattering state. Therefore, there is no reason to impose the mass-shell constraint. Because the space-time size of the fireball is of the same order as that of the nucleus, which is formed, the mismatch of the energy-momentum is washed out by the uncertainty of energy and momentum of the nucleons. 

Since the deuteron formation is a final-state process fully analogous to the generation of femtoscopic correlations, as discussed in detail in \cite{Mrowczynski:1992gc}, there is no surprise that the deuteron formation rate $A$ is given by the Sato-Yazaki formula \cite{Sato:1981ez} which is fully analogous to the Koonin formula (\ref{Koonin}) that is 
\be
\label{D-rate}
A =\frac{3}{4} (2\pi)^3 
\int d^3r_n dt_n \int d^3r_p dt_p \, S(t_n,\mathbf{r}_n) \,
S(t_p,\mathbf{r}_p) \, |\Psi_d(\mathbf{r}'_n,\mathbf{r}'_p)|^2  ,
\ee
where $\Psi_d({\bf r}_n, {\bf r}_p)$ is the deuteron wave function and $S(t,{\bf r})$, as previously, is the source function of nucleons. The neutrons and protons are assumed to be unpolarized and the spin factor $3/4$ takes into account the fact that there are 3 spin states of a spin-one deuteron and 4 spin states of a nucleon pair. If the deuteron formation rate is defined as in Eq.~(\ref{A-def}) but the there is the product of proton and neutron yields, instead of the proton yield squared, the numerical factor is $3/8$ not $3/4$ in the formula (\ref{D-rate}). The point is that a deuteron has isospin $I = 1$ while a neutron-proton pair can be in two isospin states $I = 0$ and $I = 1$.  A proton-proton pair can be only in one isospin state $I = 1$ and the extra factor $1/2$ is not needed.

Expressing the deuteron wave function with the CM variables as
\be
\Psi_d({\bf r}_n, {\bf r}_p) = e^{i{\bf P}{\bf R}} \varphi_d ({\bf r}_{np}),
\ee
the deuteron formation rate (\ref{D-rate}) equals
\be
\label{D-rate-relative}
A = \frac{3}{4} (2 \pi)^3 \int d^3r \, S^{\rm eff}_r ({\bf r}) |\varphi_d({\bf r})|^2 ,
\ee
where the relative nucleon source $S^{\rm eff}_r ({\bf r})$ is defined by Eq.~(\ref{effective-r-source}).

\subsection{Relativistic formulation}

Multiplying both sides of Eq.~(\ref{A-def}) by the nucleon energy square, one gets 
\be
\label{B-def-1}
E_d\frac{dN_d}{d^3p_d} = B \, E_n  E_p \frac{dN_{np}}{d^3 p_n\,d^3p_p} ,
\ee
where the coalescence coefficient $B$ equals
\be
\label{B-def}
B = \frac{2}{E_N} \, A = \frac{2}{m_N \, \gamma} \, A ,
\ee
$\gamma = E_N/m_N$ is the Lorentz gamma factor of the nucleon motion. Since the coefficient $B$ connects the Lorentz invariant quantities it must be Lorentz invariant as well. The coefficient depends on the deuteron momentum through $\gamma$ in the formula (\ref{B-def}) but also through the velocity ${\bf v}$ which enters the source function (\ref{effective-r-source}). So, in principle, one should write $B(p_d)$ with $p_d$ being the deuteron four-momentum. Since it is the Lorentz scalar it transforms analogously to the correlation function that is
\be
B(p_d) \longrightarrow B'(p'_d) = B (L^{-1}p'_d) .
\ee

Although the coalescence coefficient $B$ is a Lorentz scalar, it is known how to compute it only in the CM frame of the neutron-proton pair where $\gamma = 1$ and the motion of neutron and proton can be treated as non-relativistic. However, the source function needs to transformed to this reference frame. So, I compute the deuteron formation rate using the transformed source function (\ref{effective-r-source-CM}). 

\begin{figure}[t]
\centering
\includegraphics[width=11cm]{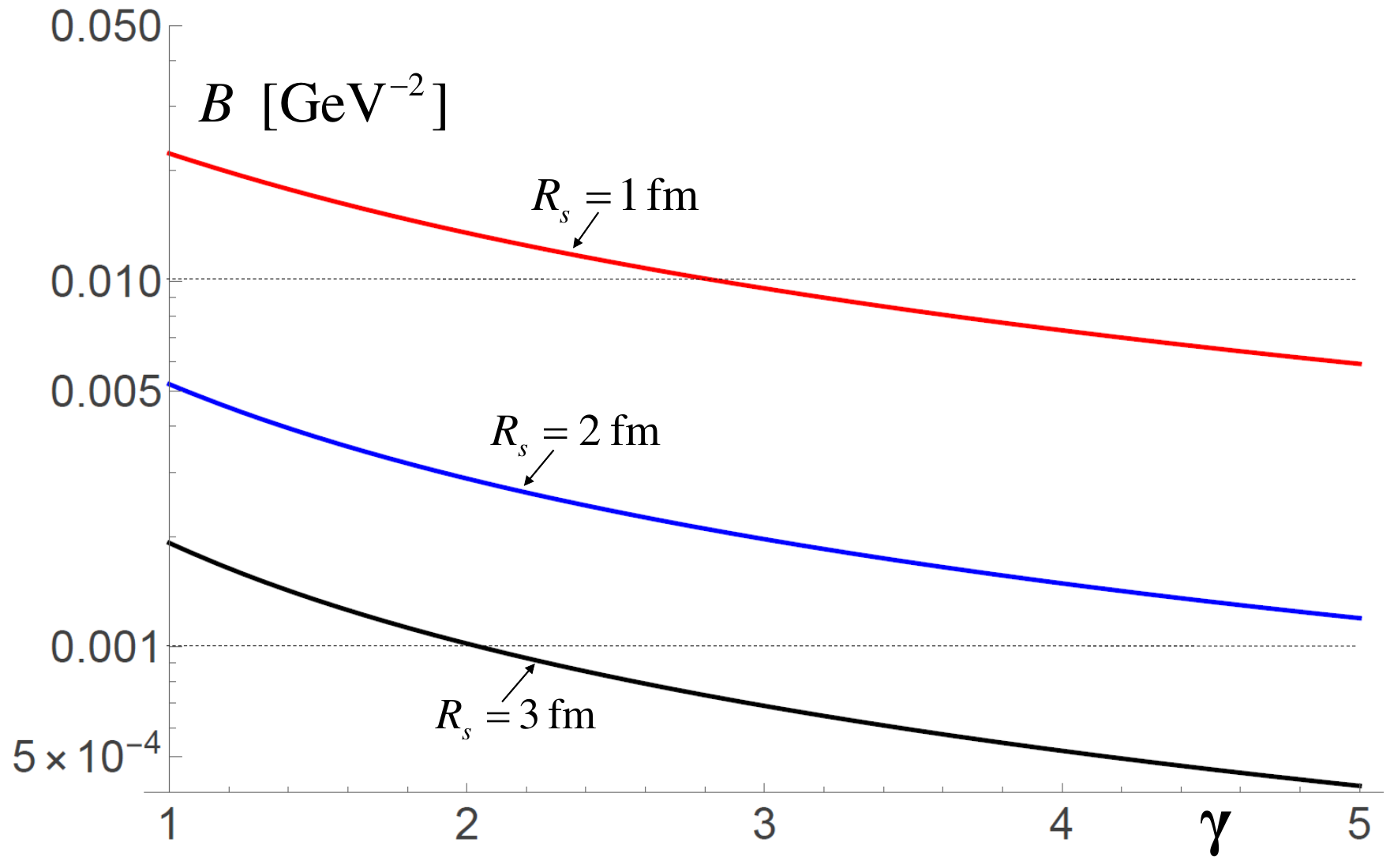}
\vspace{-3mm}
\caption{The coalescence coefficient $B$ as a function of $\gamma$.}
\label{fig-B-gamma}
\end{figure}

\subsection{Magnitude of relativistic effects}

To qualitatively understand how the coalescence coefficient $B$ depends on the velocity of the neutron-proton pair with respect to the source, I use the deuteron wave function in the Gaussian form
\be
\label{D-wave-fun-Gauss}
\varphi_d ({\bf r})  = \frac{e^{-\frac{{\bf r}^2}{8R^2_d}}}{(4 \pi R^2_d)^{3/4}} ,
\ee
where the parameter $R_d$ gives the root-mean-square radius of the deuteron as $R_{\rm rms} = \sqrt{3}\,R_d$.
Substituting the wave function (\ref{D-wave-fun-Gauss}) and source function (\ref{effective-r-source-CM}) into  Eqs.~(\ref{D-rate-relative}) and (\ref{B-def}), one finds
\be
\label{D-rate-v}
B = \frac{3}{2 m} 
\frac{ \pi^{3/2}}{\sqrt{\big(\gamma^2(R_x^2+v^2\tau^2)+ R_d^2\big)(R_y^2 + R_d^2)(R_z^2 + R_d^2)}} ,
\ee
which shows that $B$ monotonically decreases as the gamma factor grows. 

To get more reliable results one uses the deuteron wave function in the Hulth\' en form
\be
\label{Hulthen-wave}
\varphi _d({\bf r}) = \Big ( {\alpha \beta (\alpha + \beta ) \over 
2\pi (\alpha - \beta )^2} \Big )^{1/2} \;\;
{e^{-\alpha r}-e^{-\beta r}  \over r } \;, 
\ee
with $\alpha = 0.23$ fm$^{-1}$ and $\beta = 1.61$ fm$^{-1}$ \cite{Hodgson-1971}. Assuming that in the source rest frame the source function (\ref{effective-r-source-Gauss-vx}) is isotropic with $\tau = 0$ and $R_x = R_y = R_z = R_s$, the coalescence coefficient has been computed with the wave function (\ref{Hulthen-wave}). Fig.~\ref{fig-B-gamma} shows $B$ as a function of $\gamma$ for $R_s$ equal to 1, 2 and 3 fm. As one sees, $B$ decreases when $\gamma$ grows. 

The experimentally obtained coefficients $B$ show an opposite behavior: they grow with the deuteron transverse momentum both in $p$-$p$ and Pb-Pb collisions, see Fig.~3a in \cite{ALICE:2021mfm} and Fig.~5 in \cite{ALICE:2022veq}, respectively. However, the effect is strongest in the central Pb-Pb collisions, which clearly indicates that it is caused, as suggested long ago \cite{Polleri:1997bp}, by the collective radial expansion of the source. As in case of femtoscopic correlations, when the transverse momentum of the $n$-$p$ pair grows, the effective source radius in the source radius decreases and the coalescence coefficient grows. As already discussed, the source radius in the out direction includes the gamma factor, if the source radii are inferred from femtoscopic correlation functions measured in the CM frame of nucleon pairs. Then, the decrease of the source radius is even more pronounced if the source radii are given in the source rest frame. 

Since the femtoscopic $\Lambda$-$p$ and $p$-$p$ correlations provide information about the source radii of nucleon source, one can predict the coalescence coefficient $B$ using Eqs.~(\ref{D-rate-relative}) and (\ref{B-def}). Such an attempt has been undertaken in the experimental study \cite{ALICE:2021mfm}, using the theoretical formulas derived in \cite{Bellini:2020cbj} and the source radii obtained in \cite{ALICE:2020ibs}, where the source function was assumed to be isotropic in the CM frame of pairs of correlated particles. Fig.~3a in \cite{ALICE:2021mfm} shows that the coalescence coefficient $B$ computed with realistic deuteron wave functions, such as the Hulth\' en function (\ref{Hulthen-wave}), overshoots the experimental data by a factor of at least two. The agreement with the data is achieved if the coalescence coefficient is computed with the Gaussian wave function which simply predicts significantly lower $B$. However, the Gaussian function is known for long time to poorly model the deuteron wave function both at small and big distances. One asks whether the discrepancy can be resolved by taking into account the relativistic elongation of the source radius in the out direction. 

\begin{figure}[t]
\centering
\includegraphics[width=11cm]{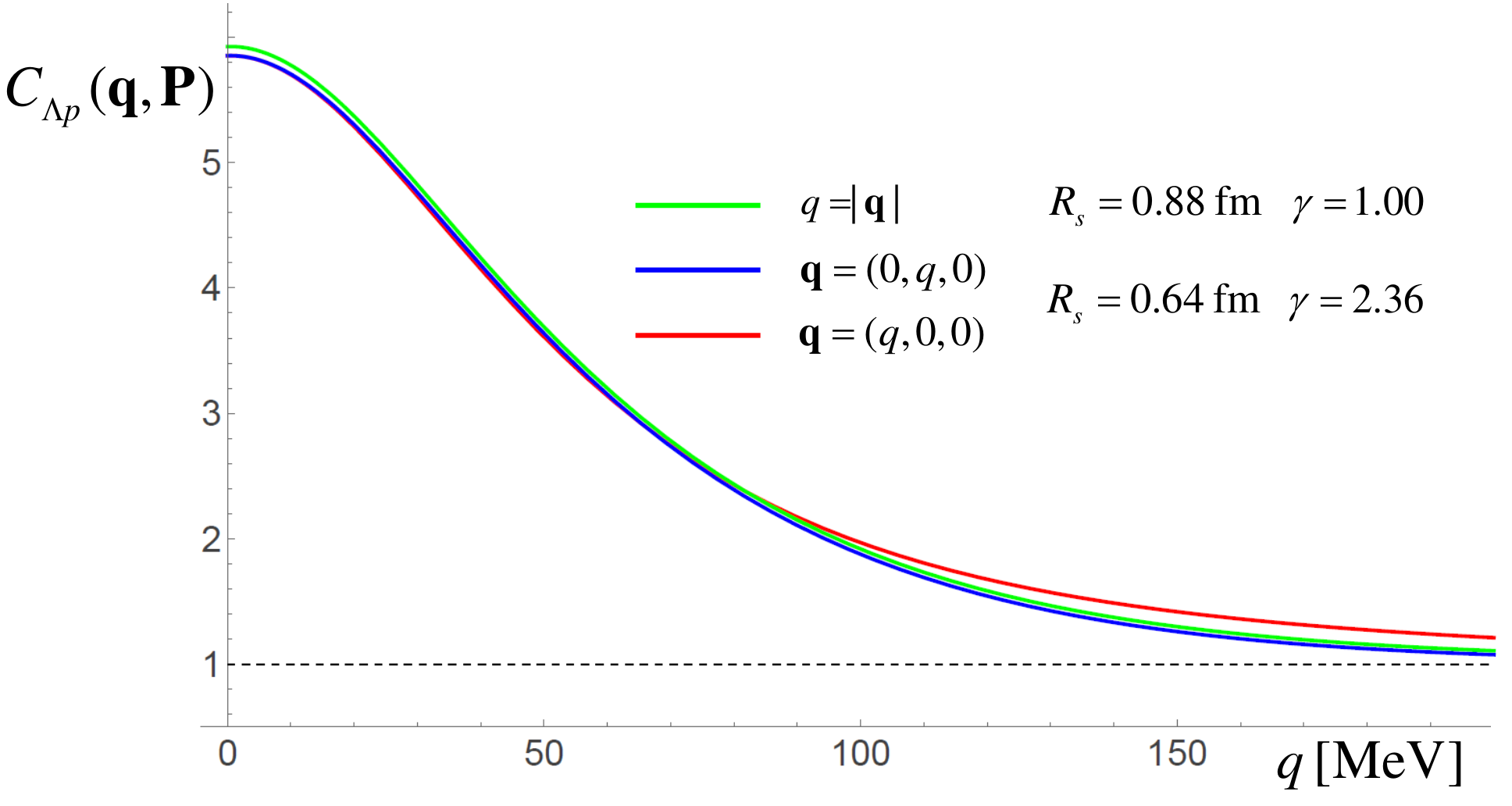}
\vspace{-3mm}
\caption{The $\Lambda$-$p$ correlation function as a function of $q =|{\bf q}|$ (green) computed for  $R_s = 0.88\,$fm and $\gamma = 1.00$ together with the $\Lambda$-$p$ correlation function as a function of  ${\bf q}=(q,0,0)$ (red) and of ${\bf q}=(0,q,0)$ computed for $R_s = 0.64\,$fm and $\gamma = 2.36$.}
\label{fig-lambda-p-iso}
\end{figure}

I consider $B$ measured at the highest transverse momentum $p_T= 2\,$GeV in the study \cite{ALICE:2021mfm}. The experimental value is $B = (1.2 \pm 0.2) \cdot 10^{-2}\,{\rm GeV}^2$. The corresponding source radius at $p_T= 2\,$GeV given in \cite{ALICE:2020ibs} is about $R_s = 0.88\,$fm and then $B = 2.9 \cdot 10^{-2}\,{\rm GeV}^2$ which overestimates the experimental value by factor bigger than 2. However, for $p_T= 2\,$GeV the gamma factor is $\gamma = 2.36$ and the source radius in the out direction is increased by this factor. So, the source function, which isotropic in the source rest frame, becomes anisotropic in the CM frame. I have checked that the $\Lambda$-$p$ correlation functions computed with $R_s = 0.88\,$fm and $\gamma = 1$ and with $R_s = 0.64\,$fm and $\gamma = 2.36$ are very close to each other, as shown in Fig.~\ref{fig-lambda-p-iso}. It would be difficult, if possible at all, to distinguish the two correlation functions experimentally. Computing the coalescence coefficient with $R_s = 0.64\,$fm and $\gamma = 2.36$, one gets $B = 2.7 \cdot 10^{-2}\,{\rm GeV}^2$ which still overshoots the experimental value. So, taking into account the relativistic effect does not resolve the problem.

\section{Summary and conclusions}
\label{sec-conclusions}

The difficulties that need to be overcome to formulate a fully relativistic approach to femtoscopic correlations and deuteron formation can be bypassed when both processes are considered in the CM frame of correlated particles or neutron-proton pairs, using methods of nonrelativistic quantum mechanics. This is justified because correlations and the probability of deuteron formation are significant only when the relative momentum of the particles of interest is small. However, working in the CM frame requires a Lorentz transformation of the source function to this frame. Contrary to a naive expectation, the source radius is not contracted but elongated in the direction of the particle's motion.

Although the calculations are performed in a preferred frame of reference, the results can be transformed to any other reference frames because the correlation function and coalescence coefficient are Lorenz scalars. The approach is not explicitly Lorentz covariant but it is covariant. It should be stressed, however, that the correlation function and coalescence coefficient must be measured at a fixed velocity of particle's pairs, as otherwise there is not enough information to perform the Lorentz transformation. 

The $\Lambda$-$p$ and $p$-$p$ correlation functions calculated with the transformed source function show that the effect of the relativistic elongation of the source radius is rather small for gamma factors $\gamma \le 4$. The correlation functions are close to isotropic. Therefore, the correlations of high-momentum particles must be measured for the relativistic effect to be clearly visible.

The traditional method to analyze the $\pi$-$\pi$ and $K$-$K$ correlations, which uses the Bowler-Sinyukov procedure to eliminate the effect of Coulomb repulsion, has been critically reviewed. I have argued that it would be desirable to use a universal methodology to study correlation functions of light and heavy hadrons. An importance of measuring the correlation functions at fixed particle pair velocities has been also emphasized. 

The coalescence factor, which has been calculated with the transformed source function, shows, in contrast to the correlation functions, a significant dependence on the Lorentz gamma factor. Taking into account the relativistic elongation of the source radius does not reconcile the description of the correlation function with the magnitude of the coalescence coefficient.

\section*{Acknowledgements}

I am grateful to Laura Fabbietti, who, during the FemTUM 2025 meeting in Sofia, asked: What about relativistic effects? This paper is an attempt to answer the question. I also thank Kfir Blum for correspondence and bringing the error in Fig.~3 to my attention.

\appendix*
\section{}

The aim of this appendix is to derive a wave function that is analog of the function  (\ref{wave-fun-asym-scatt}) when not only the short-range strong interaction but also the long-range Coulomb force is present. Due to the Coulomb interaction both the incoming plane wave and the outgoing spherical wave must be modified, as discussed in \S 138 of the textbook \cite{Landau-Lifshitz-1988}. There is also an additional complication. The wave function (\ref{wave-fun-asym-scatt}) is of the asymptotic form as the particle sources are typically bigger than or comparable to the range of the strong interaction. The source, however, is always much smaller than the Bohr radius which characterizes the length scale of the electromagnetic interaction. Therefore, one has to use the exact not asymptotic wave function which describes the electromagnetic interaction. So, one seeks a wave function analogous to (\ref{wave-fun-asym-scatt}) which is of the asymptotic form from the point of view of the strong interaction but accurately describes the electromagnetic interaction. 

In case of two non-identical particles interacting due to the repulsive Coulomb potential, the wave function can be written as
\be 
\label{coulomb-wave-2} 
\varphi_{\bf q}^C({\bf r})
= e^{i\delta_0^C} \, \sqrt{G(q)} \,  e^{i\bf qr} \,
_1F_1\Big(-{i \over a_B q}, 1, i q\eta \Big) .
\ee
The notation is explained below Eq.~(\ref{wave-function-coulomb-strong-G}).

The wave function (\ref{coulomb-wave-2}) contains both the incident and scattered waves, which can be separated at sufficiently large distances from the scattering center. It can be used to compute the correlation function that accounts only for the electromagnetic interaction; see, e.g., \cite{Maj:2009ue}. One includes the strong interaction only through the $s$-wave contribution. Therefore, because the wave function describing the $s$-wave scattering due to the combined Coulomb and strong interaction will be added separately, the $s$-wave contribution must be subtracted from the wave function (\ref{coulomb-wave-2}). This contribution reads
\be 
\label{coulomb-wave-s-final} 
\varphi_{q\, l=0}^C(r)
=   e^{i\delta_0^C} \, \sqrt{G(q)} \,  e^{iqr} \,  _1F_1\Big(1 + {i \over a_B q}, 2, - 2iqr\Big) 
= e^{i\delta_0^C} \, \frac{1}{qr} \, F_0\Big(\frac{1}{a_B q},qr\Big),
\ee
where $F_0$ is the $l=0$ regular Coulomb function (regular at $r=0$). The regular and irregular Coulomb functions $F_l$ and $G_l$ are discussed in detail in Chapter 14 of \cite{Abramowitz-Stegun-1964}.

Now, I am interested in the $l=0$ solution of the Schr\"odinger equation which takes into account the Coulomb repulsion and strong interaction. The general solution is
\be
\label{solution-F-G}
\varphi_{q\, l=0}(r) \sim \frac{1}{r} 
\bigg(F_0\Big(\frac{1}{a_B q},qr\Big) \cot\delta_0 + G_0\Big(\frac{1}{a_B q},qr\Big) \bigg),
\ee
where $G_0$ is the irregular Coulomb function, which diverges at $r=0$, and $\delta_0$ is an additional phase shift due to the strong interaction. This interpretation results from the asymptotic form of the solution (\ref{solution-F-G}) which is
\be
\varphi_{q\, l=0}(r) \sim \frac{1}{r} \, \sin\Big(qr - \frac{1}{a_B q}\log(2qr) + \delta_0^C + \delta_0 \Big) 
\ee
and holds for $qr \gg 1$. When the strong interaction is absent, the solution does not include the irregular Coulomb function. The strong interaction is expected to control a behavior of the wave function at $r \approx 0$ and thus the irregular Coulomb function is included in the general solution (\ref{solution-F-G}).

The quantity $\cot\delta_0$ is determined by postulating that the wave function and its derivative are smooth as $r \to 0$. It equals
\be
\label{cot-delta-2}
\cot\delta_0 = - \frac{2}{a_B q} \frac{h(a_B q)}{G(q)} 
+ \frac{1}{G(q)} \Big(\frac{1}{q f(q)} + i\Big) ,
\ee
where $ f(q)$ is the scattering amplitude (\ref{ampli}) due strong interaction in absence of electromagnetic one, and the function $h(x)$ is defined by Eq.~(\ref{h-def}).

Now, the incident pure Coulomb wave function (\ref{coulomb-wave-2}) with the subtracted $s$-wave contribution (\ref{coulomb-wave-s-final}) and the $s$-wave function of the strong and Coulomb interactions (\ref{solution-F-G}) are combined. Thus, one writes
\be 
\label{wave-function-coulomb-strong-Lan-Lif} 
\varphi_{\bf q}({\bf r})
= e^{i\delta_0^C} \, \sqrt{G(q)} \,  e^{i\bf qr} \,
_1F_1\Big(-{i \over a_B q}, 1, i q \eta \Big)  
-  e^{i\delta_0^C} \, \frac{F_0(r)}{qr} 
+ X\frac{1}{q r}
 \Big[G_0 \Big(\frac{1}{a_B q},qr\Big) + \cot\delta_0 F_0 \Big(\frac{1}{a_B q},qr\Big) \Big] .
\ee
The dimensionless constant $X$ should be chosen in such a way that the $s$-wave contribution (\ref{coulomb-wave-s-final}), which is subtracted from the wave function (\ref{coulomb-wave-2}), is canceled out in the wave function (\ref{wave-function-coulomb-strong-Lan-Lif}).

Keeping in mind that a scattering amplitude is expressed through the cotangent of phase shift as
\be
\label{f-0-cot}
\tilde{f}(q) = \frac{1}{q \cot \delta_0 - i q} ,
\ee
one finds the Coulomb modified amplitude (\ref{f-0}) using the formula (\ref{cot-delta-2}). Then, one shows that if
\be
X =\frac{e^{i\delta_0^C}}{q \,\tilde{f}(q)} ,
\ee
the contribution (\ref{coulomb-wave-s-final}) indeed cancels out and one obtains the wave function (\ref{wave-function-coulomb-strong-G}).

\end{document}